\documentclass[12pt]{article} 
\usepackage{bbold}
\usepackage{cite}
\usepackage{graphicx}
\def \bea{\begin{eqnarray}} 
\def \beq{\begin{equation}}

\def \eea{\end{eqnarray}} 
\def \eeq{\end{equation}}

\def \ep{\epsilon} 
\def \half{\frac{1}{2}}
\def \s{\sqrt{2}}

\def\lsim{\mathrel{\rlap{\lower3pt\hbox{$\sim$}}\raise2pt\hbox{$<$}}}
\def\gsim{\mathrel{\rlap{\lower3pt\hbox{$\sim$}}\raise2pt\hbox{$>$}}}
\def\nn{\nonumber}

\begin{document} 
\begin{flushright}
TECHNION-PH-2017-4 \\
April 2017 \\
\end{flushright} 
\centerline{\bf Reexamining the photon polarization in $B \to K\pi\pi\gamma$}
\medskip\medskip
\centerline{Michael Gronau}
\centerline{\it Physics Department, Technion, Haifa 32000, Israel}
\medskip 
\centerline{Dan Pirjol}
\centerline{National Institute of Physics and Nuclear Engineering}
\centerline{Bucharest, Romania}
\bigskip

\begin{quote}
We reexamine, update and extend a suggestion we made fifteen years ago for measuring 
the photon polarization in $b \to s\gamma$ by observing in $B \to K\pi\pi\gamma$ 
an asymmetry of the photon with respect to the $K\pi\pi$ plane. Asymmetries are 
calculated for different charged final states due to intermediate $K_1(1400)$ and 
$K_1(1270)$ resonant states. Three distinct interference mechanisms are identified 
contributing to asymmetries at different levels for these two kaon resonances. 
For $K_1(1400)$ decays including a final state $\pi^0$ an asymmetry around 
$+30\%$ is calculated, dominated by interference of two intermediate $K^*\pi$ states, 
while an asymmetry around $+10\%$ in decays including  final $\pi^+\pi^-$ 
is dominated by interference of $S$ and $D$ wave $K^*\pi$ amplitudes. In decays 
via $K_1(1270)$ to final states including a $\pi^0$ 
a negative asymmetry is favored up to $-10\%$ if one assumes $S$ wave dominance 
in decays to $K^*\pi$ and $K\rho$, while in decays involving $\pi^+\pi^-$ the
asymmetry can vary anywhere in the range $-13\%$ to $+24\%$ depending on unknown 
phases. For more precise
asymmetry predictions in the latter decays we propose studying phases in 
$K_1 \to K^*\pi, K\rho$ by performing dedicated amplitude analyses of 
$B\to J/\psi(\psi') K\pi\pi$. In order to increase statistics in studies of 
$B\to K\pi\pi\gamma$ we suggest using isospin symmetry to 
combine in the same analysis samples of charged and neutral $B$ decays.

\end{quote}

\section{Introduction}
Flavor-changing radiative $B$ meson decays provide important tests for the 
standard model. A crucial feature, which has not yet been tested experimentally 
in these processes, is the dominantly left-handed polarization of the photon in 
$b \to s\gamma$.  In several extensions of the standard model the photon in 
$b\to s\gamma$ acquire a sizable right-handed component due to chirality 
flip along a heavy fermion line in the electroweak loop process~\cite{RH}. 
A very early test for probing the dominantly left-handed photon polarization 
through time-dependent CP asymmetries, induced by interference of a large 
left-handed $b$ amplitude and a small left-handed $\bar b$ amplitude, was 
suggested in Ref.~\cite{Atwood:1997zr} and pursued experimentally by the 
Babar~\cite{Aubert:2004pe} and Belle~\cite{Ushiroda:2006fi} Collaborations.
Several years later a second test, reminiscent of a method measuring the 
tau neutrino helicity in 
$\tau \to a_1\nu_\tau, a_1 \to \rho\pi$~\cite{Kuhn:1982di,Albrecht:1990jf},
was proposed based on measuring final particle momenta in 
$B^{0,+} \to K \pi\pi\gamma$~\cite{Gronau:2001ng,Gronau:2002rz}. The 
photon polarization, a parity-odd quantity, was shown to be related to an 
asymmetry between the number of photons emitted in the two sides of the 
plane defined by $K\pi\pi$ in their center-of-mass frame.  Since this 
asymmetry is odd also under time-reversal, a potentially large asymmetry 
requires that the decay amplitude acquires a nontrivial sizable phase due to 
final state interactions. Such a large calculable phase was shown to be 
produced in $B^+\to K^0\pi^+\pi^0\gamma$ and $B^0 \to K^+\pi^-\pi^0\gamma$ 
by two interfering amplitudes involving $K^{*+}$ and $K^{*0}$ 
intermediate resonances~\cite{Gronau:2001ng,Gronau:2002rz}. 

A calculation of the decay $B \to K_1(1400)\gamma \to K\pi\pi^0\gamma$, through 
interfering amplitudes for intermediate $K^{*0}\pi$ and $K^{*+}\pi$ states, was shown 
to lead to a sizable integrated asymmetry around $34\%$~\cite{Gronau:2001ng,Gronau:2002rz}.  
The feasibility of observing such a large asymmetry in future experiments has been discussed 
in this work, assuming a branching ratio ${\cal B}(B \to K_1(1400)\gamma) = 0.7\times 10^{-5}$ 
as estimated in some models~\cite{Veseli:1995bt}. The process $B \to K_1(1270)\gamma$, 
observed a few years later with a considerably larger branching ratio [see Eqs.~(\ref{BRK1(1270)}) 
and (\ref{BRup(1400)}) below], was studied subsequently~\cite{Kou:2010kn} under 
model-dependent assumptions about the strong decay $K_1(1270) \to K \pi\pi$, thereby 
introducing a considerable uncertainty in the polarization analysis~\cite{Tayduganov:2011ui}. 
Quite recently the same authors proposed an alternative approach for obtaining this 
hadronic information by studying  the process $B \to J/\psi K_1 \to J/\psi K\pi\pi$ in 
parallel with $B \to K_1\gamma \to K\pi\pi\gamma$~\cite{Kou:2016iau}. 
A photon polarization  analysis combining contributions from several kaon resonances with 
$J^P = 1^+, 1^-, 2^+$ has been 
outlined in Ref.~\cite{Gronau:2002rz}, but would have to be treated further by 
experimental methods due to its complexity.  

The purpose of this paper is to reexamine the situation in 
$B \to K_1(1270) \gamma \to K \pi\pi\gamma$ while drawing  a comparison with 
$B \to K_1(1400)\gamma \to K\pi\pi\gamma$ which we studied only partially in
Refs.~\cite{Gronau:2001ng,Gronau:2002rz}. In contrast to Ref.\cite{Kou:2010kn} 
which applied a quark pair creation model for describing the strong decay 
$K_1(1270) \to K\pi\pi$, our approach will be purely phenomenological using 
as much information as possible from experiments. We  will discuss a few 
sources for the photon up-down asymmetry with respect to the $K_1$ decay plane, 
that are related to different types of interference occurring in $K_1$ decays. 

In Section 2 we summarize the current relevant experimental data, including 
branching ratios and certain final state interaction phases for $K_1$ decays 
to $K^*\pi$ and $\rho K$ leading to $K\pi\pi$ final states. A detailed derivation 
of relations between covariant and partial wave amplitudes describing
the latter processes is presented in Section 3 in order to resolve a discrepancy 
between relations used in Refs.~\cite{Gronau:2001ng,Gronau:2002rz} and Refs.
\cite{Kou:2010kn,Tayduganov:2011ui}. General expressions for decay amplitudes 
of $K_1 \to K\pi\pi$ are obtained in Section 4, distinguishing between hadronic  
final states involving $\pi^+\pi^-$ and $\pi^{\pm}\pi^0$. The photon up-down 
asymmetry in $B \to K\pi\pi\gamma$ with respect to the $K\pi\pi$ plane is 
calculated in Section 5 for these final states, separately for intermediate 
$K_1(1400)$ and $K_1(1270)$ resonant states. We discuss the role of three 
potential sources for an asymmetry. Section 6 uses approximate isospin 
symmetry in radiative $B$ decays to suggest combining charged and 
neutral $B\to K\pi\pi\gamma$ decays in order to increase statistics in 
studies of the photon polarization. Finally we conclude in Section 7.

\section{Experimental situation}

\subsection{$B\to K\pi\pi\gamma$}
Following the suggestions made in Refs.~\cite{Gronau:2001ng,Gronau:2002rz} for measuring the 
photon polarization in $B \to K\pi\pi\gamma$ several experiments reported measuring 
these processes. Inclusive branching ratios were measured in four charged modes, 
$B^+ \to K^+\pi^-\pi^+\gamma$, $B^0 \to K^0\pi^+\pi^-\gamma$, $B^+ \to K^0\pi^+\pi^0\gamma$
and $B^0 \to K^+\pi^-\pi^0\gamma$, for an hadronic invariant mass $m(K\pi\pi)$ in a range between 1 
GeV$/c^2$ and 1.8 or 2 Gev$/c^2$. Both the Belle~\cite{Nishida:2002me,Yang:2004as} and 
Babar~\cite{Aubert:2005xk} collaborations have observed the first two charged and neutral $B$ decay 
modes involving a pair of charged pions resulting in the following averaged branching 
ratios~\cite{Olive:2016xmw}:
\bea\label{BR1}
{\cal B}(B^+ \to K^+\pi^-\pi^+\gamma) & = & (2.76 \pm 0.22)\times 10^{-5}~,\nonumber\\
{\cal B}(B^0 \to K^0\pi^+\pi^-\gamma) & = & (1.95 \pm 0.22)\times 10^{-5}~.
\eea
Babar has also measured branching ratios for decay modes involving a neutral pion~\cite{Aubert:2005xk}:
\bea\label{BR2}
{\cal B}(B^+ \to K^0\pi^+\pi^0\gamma) & = & (4.6 \pm 0.5)\times 10^{-5}~,\nonumber\\
{\cal B}(B^0 \to K^+\pi^-\pi^0\gamma) & = & (4.1 \pm 0. 4)\times 10^{-5}~.
\eea 

Exclusive radiative $B^+$ decays involving the charged kaon resonance $K^+_1(1270)$  
decaying to $K^+\pi^-\pi^+$ have been reported by Belle~\cite{Yang:2004as},
\beq\label{BRK1(1270)}
{\cal B}(B^+ \to K_1^+(1270) \gamma)  = (4.3 \pm 1.3)\times 10^{-5}~.
\eeq
Radiative $B$ decays to $K^*_2(1430)$, first reported by the CLEO 
collaboration~\cite{Coan:1999kh},
\beq
{\cal B}(B \to K^*_2(1430)\gamma) = (1.7 \pm 0.6)\times 10^{-5}~,
\eeq
were observed subsequently by Babar at a similar rate~\cite{Aubert:2003zs}, 
\bea\label{BRK*2}
{\cal B}(B^+ \to K^{*+}_2(1430) \gamma) & = & (1.4 \pm 0.4)\times 10^{-5}~,\nonumber\\
{\cal B}(B^0 \to K^{*0}_2(1430) \gamma) & = & (1.24 \pm 0.24)\times 10^{-5}~.
\eea
We note that so far none of the $K\pi\pi\gamma$ modes observed by Belle included a 
$\pi^0$ in the final state, in contrast to several of the above measurements by Babar. 
Belle also obtained upper bounds at $90\%$ confidence level for decays involving 
$K_1(1400)$ to final states including $\pi^+\pi^-$, using only about $18\%$ of their 
final data set~\cite{Yang:2004as},
\bea\label{BRup(1400)}
{\cal B}(B^+ \to K_1^+(1400) \gamma) & < & 1.5\times 10^{-5}~,\nonumber\\
{\cal B}(B^0 \to K_1^0(1400) \gamma) & < & 1.2\times 10^{-5}~.
\eea 
These upper bounds are a factor of two larger than the branching ratio assumed in 
Ref.~\cite{Gronau:2001ng}. 

A first attempt for measuring the photon polarization in $B \to K\pi\pi\gamma$ 
was made by the LHCb collaboration~\cite{Aaij:2014wgo,Veneziano:2015ggl}. 
Nearly 14,000 signal events were reconstructed in the all charged mode 
$B^+ \to K^+\pi^-\pi^+\gamma$. The formalism developed in 
Refs.~\cite{Gronau:2001ng,Gronau:2002rz}, extended to include interference of a few kaon resonances, was applied to decay distributions for four $K\pi\pi$ mass intervals in the overall 
range $1.1 - 1.9$ GeV$/c^2$. The final result, a nonzero up-down 
asymmetry at $5.2\sigma$, was insufficient for providing a significantly quantitative 
measurement of the photon polarization.

\subsection{$K_1 \to K\pi\pi$}

An analysis of the photon polarization in $B \to K\pi\pi\gamma$ via intermediate 
$K_1(1400)$ and $K_1(1270)$ resonances requires knowledge of branching ratios 
for these kaon resonances  decaying into $K^*\pi$ and $\rho K$ states, and of magnitudes and relative phases between corresponding partial wave decay amplitudes.
The situation in decays of $K_1(1400)$ is described in Table \ref{tab.K1400}.  This information is based solely on a thirty-six-year-old experiment~\cite{Daum:1981hb} performing a partial wave analysis for $J^P=1^+$ $K\pi\pi$ states produced 
by $K^-p$ diffractive scattering with couplings to $K^*\pi$ and $\rho K$ in 
both S and D waves. In addition to measuring the ratio of $S$ and $D$ wave $K_1(1400)$ branching ratios into $K^*\pi$, some tantalizing information, 
$\delta_{DS}^{(K^*\pi)}\sim 260^\circ$, $\alpha_S \sim 40^\circ$, has been 
obtained for two relevant phases, between $K^*\pi$ $S$ and $D$ partial wave amplitudes and between $S$ wave amplitudes for $K^*\pi$ and $\rho K$, respectively.  
\begin{table}[h]
\caption{Branching fractions and particle momenta for the main decay modes of $K_1(1400)$~\cite{Olive:2016xmw}.}
\begin{center}
\begin{tabular}{c | c c c c }
\hline
Mode & ${\cal B}$ & $\Gamma_D/\Gamma_S$ & $ \delta_{DS}$ & $|\vec p|$(MeV) \\
\hline
\hline
$K^*\pi$ & $(94\pm 6)$\% & $0.04 \pm 0.01$ & $-$ & $401$\\
$\rho K$ & $(3 \pm 3)$\% & $-$ & $-$ & $291$ \\
\hline
\end{tabular}
\label{tab.K1400}
\end{center}
\end{table}
\begin{table}[h]
\caption{Branching fractions and particle momenta for the main decay modes of $K_1(1270)$~\cite{Olive:2016xmw,Guler:2010if}.}
\begin{center}
\begin{tabular}{|c|cccc|ccc|}
\hline
Mode & ${\cal B}$~\cite{Olive:2016xmw} & 
$\Gamma_D/\Gamma_S$~\cite{Olive:2016xmw} & $\delta_{DS}$ & $|\vec p|$(MeV)  
& ${\cal B}$ Fit 1~\cite{Guler:2010if} & ${\cal B}$ Fit 2~\cite{Guler:2010if} & Average \\
\hline
\hline
$\rho K$ & $(42\pm 6)$\% & $-$ & $-$ & 46  & $(57.3\pm 3.5)$\% & $(58.4\pm 4.3)$\% 
& 57.9\% \\
$K^*\pi$ & $(16\pm 5)$\% & $1.0 \pm 0.7$ & $-$ & 302 & $(26.0\pm 2.1)$\% & $(17.1\pm 2.3)$\% & 21.6\% \\
\hline
\end{tabular}
\label{tab.K1270}
\end{center}
\end{table}

The situation in decays of $K_1(1270)$ is displayed in Table \ref{tab.K1270}. The 
left-hand side is based on the same $K^- p$ scattering 
experiment~\cite{Daum:1981hb}, while the right-hand side quotes results obtained 
much more recently by the Belle Collaboration through an amplitude analysis 
determining the resonant structure of the $K^+\pi^-\pi^+$ final state in  
$B^+ \to J/\psi K^+\pi^-\pi^+$ \cite{Guler:2010if}. The difference between the $K_1(1270)$ 
decay branching ratios obtained in these two different methods seems 
to be associated with a third decay channel of $K_1(1270)$ involving 
$K^*_0(1430)\pi$, for which
a sizable branching ratio of $(28 \pm 4)\%$  was claimed in~\cite{Daum:1981hb}
in contrast to a negligible branching ratio around two percent reported 
in~\cite{Guler:2010if}. A rather crude measurement exists for the ratio of $S$ 
and $D$ wave branching ratios into $K^*\pi$~\cite{Daum:1981hb}. However no direct information exists on two relevant phases, between $K^*\pi$ partial wave amplitudes and between $S$ wave amplitudes for $K^*\pi$ and $\rho K$. A relative phase around $\phi(\rho K)-\phi(K^*\pi) \sim -40^\circ$ has been measured between total $\rho K$ and $K^*\pi$ decay amplitudes~\cite{Guler:2010if}. Assuming that these two amplitudes are dominated by an $S$ wave, this would imply 
$\alpha_S \sim -40^\circ$.

\section{Covariant and partial wave $K_1\to K^*\pi,\rho K$ amplitudes}

The amplitude for an axial-vector meson decaying to a vector meson and a 
pseudoscalar meson has two equivalent descriptions, in terms of two covariant 
amplitudes and in terms of  S and D partial wave amplitudes. The polarization 
analysis for $B \to K\pi\pi\gamma$ is based on  
covariant amplitudes~\cite{Gronau:2001ng,Gronau:2002rz} while data are given 
in terms of partial wave amplitudes. In this section we will prove relations between 
these two descriptions which will be used in our forthcoming analysis.
While these relations were given briefly in Refs.~\cite{Gronau:2001ng,Gronau:2002rz}, 
different relations have been used by the authors 
of~\cite{Kou:2010kn,Tayduganov:2011ui} quoting Ref.~\cite{Chung:1971ri} with no detail.
Here we wish to settle this discrepancy by proving these relations in some detail.

Consider, for instance $K_1 \to K^* \pi$. The covariant amplitude for 
$K_1^+(p,\ep) \to K^{*0}(p',\ep')\pi^+(p_\pi)$, involving particles with four-momenta  
$p, p', p_{\pi}$ and polarization vectors $\ep, \ep'$, is given by:
\beq\label{inv}
{\cal M}^1 = A_{K^*\pi}(\ep\cdot\ep'^*) + B_{K^*\pi}(\ep\cdot p_\pi)(\ep'^*\cdot p_\pi)~.
\eeq
In the $K_1$ rest frame ($\vec p=0$) we define $z$ as the direction of the
$K^*$ momentum, while the pion 
moves in the direction $-z$. The three possible initial spin-one $K_1$ states involving spin projection 
$\lambda=+1, 0, -1$ along $z$  are denoted $|1, \lambda\rangle$.  The three polarization vectors $\ep$ and $\ep'$ for these three states $\lambda=\pm 1,0$ 
are:
\bea\label{lam+1}
\lambda=1 &:&  \ep = \ep' = (0, -\sqrt{\half}(\vec e_1 + i\vec e_2))~,
\\
\label{lam-1}
\lambda =-1 &:& \ep = \ep' = (0, \sqrt{\half}(\vec e_1 - i\vec e_2))~,
\\
\lambda=0 &:& \ep = (0, \vec e_3), \ep' = (|\vec p_\pi|/m_{K^*}, 
(E_{K^*}/m_{K^*})\vec e_3)~.
\eea
For $\lambda=1$ this is the form of $\ep$ in the $K_1$ rest frame. The same form in this frame, identical to its form in the $K^*$ rest frame, applies to $\ep'$ because a Lorentz transformation along $z$ does not change the $x, y$ components, mixing 
only the $t, z$ components. For $\lambda=0$ $\ep'$ is obtained from $\ep'(K^*) = (0, \vec e_3)$ in the rest frame of $K^*$  by a Lorentz transformation to the rest frame of $K_1$ using $\gamma = E_{K^*}/m_{K^*}, \gamma\beta = -|p_\pi|/m_{K^*}$,
\bea
\ep'_0(K_1) & = & \gamma[\ep'_0(K^*) - \beta\ep'_3(K^*)] = |\vec p_\pi|/m_{K^*}~,
\nonumber\\
\ep'_3(K_1) & = & \gamma[\ep'_3(K^*) - \beta \ep'_0(K^*)] = E_{K^*}/m_{K^*}~.
\eea 
We note that the transversity condition $\ep' \cdot p'=0$ is satisfied for 
all polarization states $\lambda$ of the $K^*$ meson. 
In particular, for $\lambda=0$  we have,
using $p'(K^*) = (E_{K^*}, |\vec p_\pi|\vec e_3)$,
\beq
p'(K^*)\cdot \ep'(K^*) \hskip-1mm= \hskip-1mm 
   E_{K^*}|\vec p_\pi|/m_{K^*} - |\vec p_\pi|E_{K^*}/m_{K^*} =0~.
\eeq

The covariant decay amplitude (\ref{inv}) can now be calculated for these three polarization states:
\bea\label{Mpm1}
\lambda & = & \pm 1: (\ep\cdot\ep'^*) = -\half(\vec e_1 \pm i\vec e_2)(\vec e_1 
\mp i\vec e_2) = -1;~\ep\cdot p_\pi = 0
\nonumber\\
&& \Rightarrow  {\cal M}^1_{\lambda =\pm 1} = -A_{K^*\pi}~.
\eea
\bea\label{M0}
&& \lambda =0: \ep\cdot\ep'^* = -\frac{E_{K^*}}{m_{K^*}};~\ep\cdot p_\pi 
= |\vec p_\pi|; \ep'^*\cdot p_\pi = \frac{E_{\pi}|\vec p_\pi|}{m_{K^*}} + 
\frac{E_{K^*}|\vec p_\pi|}{m_{K^*}} = \frac{m_{K_1}|\vec p_\pi|}{m_{K^*}}
\nonumber\\
&& \Rightarrow   {\cal M}^1_{\lambda = 0} = - A_{K^*\pi}\frac{E_{K^*}}{m_{K^*}} + 
B_{K^*\pi}\frac{m_{K_1}|\vec p_\pi|^2}{m_{K^*}}~.
\eea

Let us now write decay amplitudes for the three polarization states 
$|1, \lambda\rangle$ in terms of amplitudes for S and D  waves, 
$L=0, 2$, noting that the angular momentum states carry $L_z =0$ ($m=0$) for 
$K^*$ and $\pi$ moving in $\pm z$ directions. 
Using SU(2) Clebsch-Gordan coefficients $(l~0; 1~\lambda | 1~\lambda)$ and 
absorbing a factor $1/\sqrt{5}$ in the definition of the D-wave amplitude, 
we have: 
\beq\label{SDpm1}
{\cal M}^1_{\lambda = \pm 1} = (0~0 ; 1 \pm 1 | 1 \pm 1)C^{(K^*\pi)}_S + (2~0; 1 \pm 1|1 \pm 1)\sqrt{5}C^{(K^*\pi)}_D = C^{(K^*\pi)}_S + \frac{1}{\sqrt{2}}\,C^{(K^*\pi)}_D~,
\eeq
\beq\label{SD0}
{\cal M}^1_{\lambda = 0} = (0~0 ; 1~0 | 1~0)C^{(K^*\pi)}_S + 
(2~0; 1~0|1~0)\sqrt{5}C^{(K^*\pi)}_D = C^{(K^*\pi)}_S  -
\s\,C^{(K^*)}_D~.
\eeq
Squaring magnitudes of these amplitudes and averaging over 
the three polarizations states of  the $K_1$ meson, one obtains
\beq
\frac{1}{3}\sum_{\lambda=0,\pm 1}
 |{\cal M}^1_\lambda|^2 = |C^{(K^*\pi)}_S|^2 + 
|C^{(K^*\pi)}_D|^2~,
\eeq
implying a decay rate 
\beq\label{rate}
\Gamma(K_1 \to K^*\pi)  = \frac{1}{8\pi m^2_{K_1}}
(|C^{(K^*\pi)}_S|^2 + |C^{(K^*\pi)}_D|^2) |\vec p_\pi|\,.
\eeq

Comparing Eqs.~(\ref{Mpm1}) and (\ref{M0}) with (\ref{SDpm1}) and (\ref{SD0}) one obtains
\bea\label{ASD}
-A_{K^*\pi} & = & C^{(K^*\pi)}_S + \frac{1}{\sqrt{2}}C^{(K^*\pi)}_D~,
\\
-B_{K^*\pi}\frac{m_{K_1}|\vec p_\pi|^2}{m_{K^*}} & = & 
C^{(K^*\pi)}_S\left(\frac{E_{K^*}}{m_{K^*}} - 1\right) + \frac{1}{\sqrt{2}}C^{(K^*\pi)}_D\left(\frac{E_{K^*}}{m_{K^*}} + 2\right)~,
\eea
or
\beq\label{BSD}
-B_{K^*\pi}|\vec p_\pi|^2 = C^{(K^*\pi)}_S\left(\frac{E_{K^*}-m_{K^*}}{m_{K_!}}\right)
+ \frac{1}{\sqrt{2}}C^{(K^*\pi)}_D\left(\frac{E_{K^*} + 2m_{K^*}}{m_{K_1}} \right)~.
\eeq 
These relations agree with those applied in 
Refs.~\cite{Gronau:2001ng,Gronau:2002rz} using a different convention for 
partial wave amplitudes. [The amplitudes $C_{S,D}^{(K^*\pi)}$ are related to $c_{S,D}$ occurring in Eq.~(20) of \cite{Gronau:2002rz} by $C_S^{(K^*\pi)} = - c_S$ and 
$C_D ^{(K^*\pi)} = - \sqrt2 \frac{m_{K_1} |\vec p_\pi|^2}{2m_{K^*}+E_{K^*}}c_D$.]

While the expression for $A_{K^*\pi}$ agrees with the one quoted by the authors of 
Ref.~\cite{Kou:2010kn}, these authors used a different relation for 
$B_{K^*\pi}$. Their Eq.~(27) reads in our notation~\cite{eq27},
\beq\label{Bkou}
-B_{K^*\pi}|\vec p_\pi|^2 = \frac{E_{K^*}}{m_{K^*}}
\left[C^{(K^*\pi)}_S\left(\frac{m_{K^*}-E_{K^*}}{m_{K_1}}\right) 
+ \frac{1}{\sqrt{2}}C^{(K^*\pi)}_D\left(\frac{m_{K^*} + 2E_{K^*}}{m_{K_1}}\right)\right]~.
\eeq
We find that this relation is in disagreement with (\ref{BSD}), 
and is therefore incorrect.

Relations similar to (\ref{ASD}) and (\ref{BSD}) apply to $K_1 \to K\rho$:
\bea\label{ASDr}
-A_{K\rho} & = & C^{(K\rho)}_S + \frac{1}{\sqrt{2}}C^{(K\rho)}_D~,
\\\label{BSDr}
-B_{K\rho}|\vec p_K|^2 & = & C^{(K\rho)}_S\left(\frac{E_{\rho}-m_{\rho}}{m_{K_!}}\right)
+ \frac{1}{\sqrt{2}}C^{(K\rho)}_D\left(\frac{E_{\rho} + 2m_{\rho}}{m_{K_1}} \right)~.
\eea
\section{Decay amplitudes for $K_1 \to K\pi\pi$} 

The two pairs of processes in Eqs.~(\ref{BR1}) and (\ref{BR2}) obtain 
contributions from $K_1(1270)$ and $K_1(1400)$ resonances. The decays of 
these resonances to different $K\pi\pi$ charged modes may be divided into two 
distinct pairs distinguished by their intermediate resonant decay 
channels~\cite{Gronau:2002rz}. 
The first pair involves two single decay channels into $K^*\pi$ and $K\rho$, 
\beq
K^+_1 \to \left\{ \begin{array}{c }  K^{*0}\pi^+ \cr
K^+ \rho^0\end{array} \right\} \to K^+\pi^-\pi^+~,
\eeq
\beq
K^0_1 \to \left\{ \begin{array}{c }  K^{*+}\pi^- \cr
K^0 \rho^0 \end{array} \right\} \to K^0\pi^+\pi^-~,
\eeq
while the second pair obtains contributions from two interfering $K^*\pi$ decay 
channels in addition to a single $K\rho$ channel,
\beq
K^+_1 \to \left\{ \begin{array}{c }  K^{*+}\pi^0 \cr
K^{*0} \pi^+ \cr K^0 \rho^+
\end{array} \right\} \to K^0\pi^+\pi^0~,
\eeq
\beq
K^0_1 \to \left\{ \begin{array}{c }  K^{*+}\pi^- \cr
K^{*0}\pi^0 \cr K^+\rho^- \end{array} \right\} \to K^+\pi^-\pi^0~.
\eeq

The decays of $K_1^+$ and $K_1^0$ within each pair are related to each 
other by isospin reflection $u \leftrightarrow d$, implying equal amplitudes 
in the isospin symmetry limit:
\bea\label{type3}
A(K^+_1 \to K^+\pi^-\pi^+) & = & A(K^0_1 \to K^0\pi^+\pi^-)~,
\\\label{type2}
A(K^+_1 \to K^0\pi^+\pi^0) & = &  A(K^0_1 \to K^+\pi^-\pi^0)~.
\eea
These two amplitudes, for final states characterized by two charged pions in one case 
and by a pair of charged and neutral pions in the other, will be studied separately. 
Only the second pair of amplitudes has been analyzed for $K_1(1400)$ in 
Refs.~\cite{Gronau:2001ng,Gronau:2002rz}.

\subsection{Decays involving two charged pions $K_1^+ \to K^+\pi^+\pi^-,
K_1^0\to K^0\pi^+\pi^-$} 

The $K^{*0}\pi^+$ contribution to the decay amplitude for 
$K_1^+(p,\ep) \to K^+(p_3)\pi^+(p_1)\pi^-(p_2)$ is obtained by
convoluting the amplitude (\ref{inv}) with the amplitude for $K^{*0} \to K^+\pi^-$, 
\beq
A(K^{*0} \to K^+ \pi^-) = g_{_{K^*K\pi}}\ep'\cdot (p_2 - p_3)~,
\eeq
including a Breit-Wigner propagator for the $K^*$,
\bea
&& \left[g^{\mu\nu}  - \frac{(p_2 + p_3)^{\mu}(p_2 + p_3)^{\nu}}{m^2_{K^*}}\right]
{\cal B}^{(K^*)}_{23}~,
\nonumber\\
&& {\cal B}^{(K^*)}_{23} \equiv(s_{23} - m^2_{K^*} - im_{K^*}\Gamma_{K^*})^{-1}~,~~~
s_{23} \equiv (p_2 + p_3)^2~.
\eea
A similar contribution due to $K_1^+ \to  K^+\rho^0, \rho^0\to \pi^+\pi^-$ 
involves 
invariant amplitudes $A_{\rho K}, B_{\rho K}$ describing $K_1 \to  K\rho$, the 
strong coupling 
$g_{\rho\pi\pi}$ and a Breit-Wigner propagator for the $\rho$, 
${\cal B}^{(\rho)}_{12} \equiv (s_{12} - m^2_\rho - im_\rho\Gamma_\rho)^{-1}$. 
Specifically, we define the amplitude 
\beq
A(\rho^0 \to \pi^+ \pi^-) = g_{_{\rho\pi\pi}}\ep'\cdot (p_1 - p_2)~.
\eeq

Adding these two contributions and neglecting a non-resonant term 
(which is justified
in $K_1(1400)$ more than in $K_1(1270)$ - see Tables \ref{tab.K1400} and \ref{tab.K1270}), 
the total covariant amplitude for $K_1^+(p,\ep) \to K^+(p_3)\pi^+(p_1)\pi^-(p_2)$
is given by
\beq\label{calM}
{\cal M} = C_1(\ep\cdot p_1) - C_2(\ep\cdot p_2)~,
\eeq
where 
\beq
C_i = C_i^{(K^*\pi)} + C_i^{(K\rho)}~~(i=1, 2)~,
\eeq
\bea\label{Ci}
& C_1^{(K^*\pi)} = g_{_{K^*K\pi}}{\cal B}^{(K^*)}_{23}\{A_{K^*\pi} + 
B_{K^*\pi}\,p_1\cdot(p_2-p_3)
- \frac{m^2_K - m^2_\pi}{m^2_{K^*}}[A_{K^*\pi} - 
B_{K^*\pi}((p\cdot p_1) - m^2_\pi)]\}~,
\nonumber\\
& C_1^{(K\rho)} =  g_{\rho\pi\pi}{\cal B}^{(\rho)}_{12}[A_{K\rho} - 
B_{K\rho}(p\cdot p_1 - p\cdot p_2)]~,
\nonumber\\ 
& C_2^{(K^*\pi)}  = -g_{_{K^*K\pi}}{\cal B}^{(K^*)}_{23}(2A_{K^*\pi})~,
\nonumber\\ 
& C_2^{(K\rho)} = - g_{\rho\pi\pi}{\cal B}^{(\rho)}_{12}[A_{K\rho} + 
B_{K\rho}(p\cdot p_1 - p\cdot p_2)]~.
\eea

The four scalar products of two momenta in (\ref{Ci}), $p_1\cdot p_2$, 
$p_1\cdot p_3$, 
$p \cdot p_1$ and $p\cdot p_2$, may all be written in terms of 
$s_{13}$ and $s_{23}$. That is, $C_i$ are functions of these two variables and
the decay amplitude has the explicitly covariant form
\beq
{\cal M} = C_1(s_{13}, s_{23})(\ep\cdot p_1) - C_2(s_{13}, s_{23})(\ep\cdot p_2)~.
\eeq

\subsection{Decays involving a neutral pion $K_1^+ \to K^0\pi^+\pi^0,
K_1^0\to K^+\pi^-\pi^0$}

In these decays the amplitude has the same structure as (\ref{calM}),
\beq\label{M'}
{\cal M'}  =  C'_1(s_{13},s_{23})(\ep\cdot p_1) - C'_2(s_{13}, S_{23})(\ep\cdot p_2)~,
\eeq
with two contributions to $C'_{1,2}$ from $K^{*0}\pi$ and $K^{*+}\pi$ and one contribution from $K\rho^{\pm}$. 
The overall contribution from $K^*\pi$ is antisymmetric under the exchange of 
the two pion momenta and, using isospin, is expressed in terms of the same quantities 
$C_i^{(K^*\pi)}$ given in (\ref{Ci}):
\beq\label{C'}
C_i^{\prime(K^*\pi)} = \sqrt{\half}[C^{(K^*\pi)}_i - C^{(K^*\pi)}_i (p_1 \leftrightarrow p_2)]~.
\eeq
The single contribution from $K\rho$ is
\beq
C_i^{\prime(K\rho)} = \s C_i^{(K\rho)}~.
\eeq

\subsection{Experimental information on ratios of amplitude}

In the next section studying the photon polarization in $B \to K_1\gamma, 
K_1 \to K\pi\pi$, which depends on interference of amplitudes, we will need 
ratios of certain quantities which we calculate now.

The strong couplings $g_{_{K^*K\pi}}$ and $g_{\rho\pi\pi}$ occurring in 
(\ref{Ci}) (for which we used a slightly different convention 
in~\cite{Gronau:2002rz}) are obtained from the $K^*$ and $\rho$ widths. Using
\bea
&& \Gamma(K^{*0}\to K^+\pi^-) =  \frac23 \Gamma_{K^*}\,\mathcal{B}(K^*\to K\pi) 
= \frac{1}{6\pi m_{K^*}^2} |g_{K^* K\pi}|^2
|\vec p_\pi|^3~, \\
&& \Gamma(\rho^0 \to \pi^+\pi^-) =  \Gamma_{\rho}\,\mathcal{B}(\rho\to \pi\pi) 
=  \frac{1}{6\pi m_{\rho}^2} |g_{\rho\pi\pi}|^2|\vec p_\pi|^3~,
\eea
where $\Gamma_{K^*} = 51$ MeV, $\Gamma_{\rho} = 150$ MeV, 
$|\vec p_\pi|_{K^*\to K\pi} = 289$ MeV,  $|\vec p_\pi|_{\rho\to \pi\pi} = 364$ MeV
\cite{Olive:2016xmw}, we calculate
\beq\label{gratio}
\frac{|g_{\rho\pi\pi}|}{|g_{_{K^*K\pi}}|} = 1.29 \,.
\eeq
This compares well with an SU(3) prediction
\beq
\frac{g_{\rho\pi\pi}}{g_{_{K^*K\pi}}} = - \sqrt2 \,.
\eeq

The quantities $A_{K^*\pi}, B_{K^*\pi}, A_{K\rho}, B_{K\rho}$ in Eqs.~(\ref{Ci}) 
may be obtained from $S$ and $D$ wave amplitudes measured in 
$K_1^+ \to K^{*0}\pi^+$ and $K_1^+\to K^+\rho^0$ decays, denoted 
$C^{(K^*\pi)}_{S,D}$ and $C^{(K\rho)}_{S,D}$, using Eqs.~(\ref{ASD}) (\ref{BSD}) 
for the first process and (\ref{ASDr}) (\ref{BSDr}) for the second. 
Branching ratios for $K_1 \to K^*\pi$ and $K_1 \to K\rho$ summed over all charged 
modes and corresponding ratios of decay rates for $S$ and $D$ waves waves were 
given in Tables \ref{tab.K1400} and \ref{tab.K1270} for $K_1(1400)$ and $K_1(1270)$, 
respectively. We will denote by $\delta_{DS}^{(K^*\pi)}$ and 
$\delta_{DS}^{(K\rho)}$ relative phases between $S$ and $D$ wave 
amplitudes in $K_1^+ \to K^{*0}\pi^+$ and $K_1^+ \to K^+\rho^0$, respectively, and by 
$\kappa_S$ and $\alpha_S$ the magnitude and phase of the ratio of $S$ wave 
amplitudes for these decays,
\beq
\delta_{DS}^{(K^*\pi)} \equiv {\rm arg}(C_D^{(K^*\pi)}/C_S^{(K^*\pi)})\,,~~~~
\delta_{DS}^{(K\rho)} \equiv {\rm arg}(C_D^{(K\rho)}/C_S^{(K\rho)})\,,~~~
\kappa_S e^{i\alpha_S} \equiv C_S^{(K\rho)}/C_S^{(K^*\pi)}\,.
\eeq

Ratios of amplitude will now be calculated separately for $K_1(1400)$ and 
$K_1(1270)$ applying Eqs.~(\ref{ASD}) (\ref{BSD}) 
to $K_1^+ \to K^{*0}\pi^+$ and (\ref{ASDr}) (\ref{BSDr}) to $K_1^+ \to K^+ \rho^0$.
Meson masses will be taken from~\cite{Olive:2016xmw}.
\begin{itemize}
\item $K_1(1400)$
\vskip 1mm
Using $|C_D^{(K^*\pi)}|^2/|C_S^{(K^*\pi)}|^2= 0.04 \pm 0.01$ and since 
the branching ratio $K_1 \to \rho K$  is very small, we calculate:
\bea\label{ratio1}
\frac{B_{K^*\pi}}{A_{K^*\pi}} & = & 
\frac{0.38 + 1.73 e^{i\delta_{DS}^{(K^*\pi)}}}
{1 + 0.14 e^{i\delta_{DS}^{(K^*\pi)}}}~,
\\
\label{ratio2}
\frac{B_{K\rho}}{A_{K\rho} }& = &
\frac{0.45 + C_D^{(K\rho)}/C_S^{(K\rho)}}
{1+ 0.05 C_D^{(K\rho)}/C_S^{(K\rho)}} \sim 0.45~,
\eea
The ratio $|C_S^{(K\rho)}|/|C_S^{(K^*\pi)}|$ may be obtained from
\beq\label{45}
\frac{{\cal B}(K_1\to [K\rho]_S)}{{\cal B}(K_1 \to [K^*\pi]_S)} = 
\frac{2|C_S^{(K\rho)}|^2|\vec p_K|}{|C_S^{(K^*\pi)}|^2|\vec p_\pi|}~,
\eeq
implying together with (\ref{gratio}) and assuming a central value 
${\cal B}(K_1 \to K\rho) = 3\%$, 
\bea
\label{ratio3}
\kappa_S \equiv \frac{g_{\rho\pi\pi}}{g_{_{K^*K\pi}}}
\frac{C_S^{(K\rho)}}{C_S^{(K^*\pi)}} & = &0.19 e^{i\alpha_S}~.
\eea
[The relative phase $\alpha_S$ between the $[K^*\pi]_S$ and $[\rho K]_S$ 
amplitudes was quoted as $20^\circ < \alpha_S < 60^\circ$ in 
\cite{Gronau:2001ng,Gronau:2002rz}, following the ACCMOR paper 
\cite{Daum:1981hb}.]
The factor of 2 on the right-hand side of (\ref{45}) is due to the
specific choice of the modes $K_1^+\to K^{*0}\pi^+$ and $K_1^+\to \rho^0 K^+$
used to define the couplings $C_S^{(K^*\pi)}$ and $C_S^{(\rho K)}$,
while the branching ratios on the left-hand side are for final states summed over 
all charges.
\item $K_1(1270)$
\vskip 1mm
Taking the central value in $|C_D^{(K^*\pi)}|^2/|C_S^{(K^*\pi)}|^2= 1.0 \pm 0.7$
 and assuming that $S$ wave dominates $K_1 \to \rho K$ because of an extremely 
 small available phase space, we find:
\bea
\frac{B_{K^*\pi}}{A_{K^*\pi}} & = & 
\frac{0.43 + 16.6 e^{i\delta_{DS}^{(K^*\pi)}}}
{1 + 0.71 e^{i\delta_{DS}^{(K^*\pi)}}}~,
\\
\frac{B_{K\rho}}{A_{K\rho} }& = & 0.51~.
\eea
Using for branching ratios of $K_1 \to K^*\pi$ and $K_1\to K\rho$ 
averages of the two Belle fits in Table~\ref{tab.K1270}, we calculate
\bea
\kappa_S \equiv \frac{g_{\rho\pi\pi}}{g_{_{K^*K\pi}}}
\frac{C_S^{(K\rho)}}{C_S^{(K^*\pi)}} & = & 5.42 e^{i\alpha_S} \,.
\eea
A relative phase $\phi(\rho K)- \phi(K^*\pi) \sim -40^\circ$ between total
amplitudes has been measured by the Belle collaboration~\cite{Guler:2010if}.
However, translating this into a constraint on $\alpha_S$
requires information about the partial wave amplitudes which is not available.
\end{itemize}
  
 \section{Photon polarization and asymmetry in $B\to K\pi\pi\gamma$}

We have shown that the decay amplitude for 
$K_1(\vec p=0,\vec\ep) \to \pi(p_1)\pi(p_2)K(p_3)$ in the $K_1$ rest frame has the 
general structure
\beq
{\cal M}_{\rm rest} = C_1(s_{13}, s_{23})\vec p_1\cdot\vec\ep - 
C_2(s_{13}, s_{23})\vec p_2\cdot\vec\ep = \vec J\cdot \vec\ep~,
\eeq
where
\beq\label{J}
\vec J \equiv C_1(s_{13}, s_{23})\vec p_1 - C_2(s_{13}, s_{23})\vec p_2~.
\eeq
Considering now $B\to K_1\gamma$ followed by $K_1 \to K\pi\pi$ we wish 
to study the angular distribution of the photon with respect to the $K_1$ decay plane 
as function of the photon polarization. For completeness  we will derive this relation,
although some parts of the derivation can be found in 
Refs.~\cite{Gronau:2001ng,Gronau:2002rz}. One reason for presenting this complete analysis is correcting a sign error in defining a specific direction in this previous work.

Working in  the rest frame of $K_1$, we take the photon momentum
$\vec p_\gamma$ along the $-z$ direction, and the $B$ meson momentum
along the $+z$ direction. There are two amplitudes for $\bar B\to K_1\gamma$
decays, corresponding to left- and right-handed photons
\begin{eqnarray}
\mathcal{M}_L \equiv \mathcal{A}(\bar B \to K_1 \gamma_L) \,,\quad
\mathcal{M}_R \equiv \mathcal{A}(\bar B \to K_1 \gamma_R)~.
\end{eqnarray}
Defining the photon polarization parameter,
\begin{eqnarray}
\lambda_\gamma \equiv \frac{|\mathcal{M}_R|^2 -|\mathcal{M}_L|^2}
{|\mathcal{M}_R|^2 +|\mathcal{M}_L|^2}\,,
\end{eqnarray}
we would like to determine $\lambda_\gamma$ through the angular distribution 
of the decay products of the $K_1$ meson. 

The amplitude for $\bar B \to K\pi\pi\gamma_{L,R}$ is proportional to the 
decay amplitude $K_1(p,\epsilon)\to K\pi\pi$ with $\epsilon=\epsilon_{L,R}$
corresponding to the transverse polarization states $|\lambda = \mp 1\rangle$ 
of the $K_1$ meson in its rest frame [see Eqs.\,(\ref{lam+1}) (\ref{lam-1})],
\bea
\mathcal{A}(\bar B \to K\pi\pi \gamma_{L,R}) = \mathcal{M}_{L,R}
(\vec \epsilon_{\mp 1}\cdot \vec J) = 
\pm \frac{1}{\sqrt2} \mathcal{M}_{L,R} (J_x \mp i J_y) \,.
\eea
Squaring the amplitude,
\bea
|\mathcal{A}(\bar B \to K\pi\pi \gamma_{L,R})|^2 &=& 
\frac12|\mathcal{M}_{L,R}|^2
(J_x \mp i J_y)(J_x^* \pm i J_y^*) 
\nonumber\\
&=& \frac12|\mathcal{M}_{L,R}|^2
\left\{ |J_x|^2 + |J_y|^2 \mp 2\mbox{Im}(J_x J_y^*) \right\}\,,
\eea
and summing  over the two photon polarization states, one obtains
\bea\label{avsquare}
&& \sum_{\lambda=L,R}|\mathcal{A}(\bar B \to K\pi\pi \gamma_{\lambda})|^2 
\nonumber\\
&& = \frac12 
\{ |\mathcal{M}_L|^2 + |\mathcal{M}_R|^2 \} \left[( |J_x|^2 + |J_y|^2)
+ 2\lambda_\gamma \mbox{Im}(J_x J_y^*) \right]~. 
\eea

We denote by $\hat n = (\vec p_1\times \vec p_2)/|\vec p_1\times \vec p_2|$ 
the normal to the $K_1$ decay plane defined by the
two pions momenta. 
The orientation of the $K\pi\pi$ plane with respect to the $(\hat e_x,\hat e_y,
\hat e_z)$ axes is determined by three Euler-like angles $(\theta,\phi,\psi)$. 
The polar angles $(\theta, \psi)$ define the orientation of $\hat n$ with
respect to $\hat e_z$ such that $\cos\theta = \hat e_z \cdot \hat n$, and the 
third angle $\phi$ parameterizes rotations of the $K\pi\pi$
plane around $\hat n$. The intersection of the 
$K\pi\pi$ plane with the $(\hat e_x, \hat e_y)$ plane is the nodal line,
and its angle with respect to $\hat e_x$ is $\psi$. We denote unit vectors 
in the $K\pi\pi$ plane by $(\hat e_1,\hat e_2)$ such that 
$\hat e_3 \equiv \hat n$,  and define $\phi$ as the angle between the 
nodal line and $\hat e_1$.

The vector $\vec J$ lies in the $(\hat e_1, \hat e_2)$ plane. Its components
in the $(\hat e_x,\hat e_y,\hat e_z)$ coordinates can be expressed in terms
of the angles introduced above:
\begin{eqnarray}\label{Jxyz}
&& J_x = (J_1 \cos\phi + J_2 \sin\phi) \cos\psi 
       - (- J_1 \sin\phi + J_2 \cos\phi) \sin\psi\cos\theta~, \nn \\
&& J_y = (J_1 \cos\phi + J_2 \sin\phi) \sin\psi 
       + (- J_1 \sin\phi + J_2 \cos\phi) \cos\psi\cos\theta~, \nn \\
&& J_z = (-J_1 \sin\phi + J_2 \cos\phi) \sin\theta~.
\end{eqnarray}
These equations are obtained by noting that the projections of $\vec J$ in the
$K\pi\pi$ plane,  $J_\parallel$ along the nodal line and $J_\perp$ perpendicular 
to it, are
\begin{eqnarray}\label{Jpp}
J_\parallel = J_1 \cos\phi + J_2 \sin\phi \,, \quad
J_\perp = - J_1 \sin\phi + J_2 \cos\phi\,.
\end{eqnarray}
The components along the $(\hat e_x,\hat e_y)$ directions are
\begin{eqnarray}
&& J_x = J_\parallel \cos\psi - J_\perp \sin \psi \cos\theta~, \\
&& J_y = J_\parallel \sin\psi + J_\perp \cos \psi \cos\theta~.
\end{eqnarray}
Substituting (\ref{Jpp}) in these relations leads to (\ref{Jxyz}).

Using
\beq
|J_x|^2 + |J_y|^2 = |\vec J|^2 - |J_z|^2~,
\eeq
and averaging over $\phi$ implies for the first term in (\ref{avsquare}), 
\begin{eqnarray}
\frac{1}{2\pi} \int_0^{2\pi} d\phi (|\vec J_x|^2 + |J_y|^2) = 
|\vec J|^2 ( 1 - \frac12 \sin^2\theta) = \frac12 |\vec J|^2 (1 + \cos^2\theta)~.
\end{eqnarray}
The second term multiplying $\lambda_\gamma$ is
\begin{eqnarray}
\mbox{Im}(J_x J_y^*) &=& \mbox{Im}(J_\parallel J_\perp^*) \cos^2\psi \cos\theta
- \mbox{Im}(J_\parallel^* J_\perp) \sin^2\psi \cos \theta \\
&=& 2 \mbox{Im}(J_\parallel J_\perp^*) \cos\theta =
2 \mbox{Im}(J_1 J_2^*) \cos\theta =
\mbox{Im}[\hat n \cdot (\vec J \times \vec J^*)] \cos\theta 
\nn \,.
\end{eqnarray}
Thus, after averaging over rotations in the $K\pi\pi$ decay plane
(angle $\phi$) and around the $\hat e_z$ axis (angle $\psi$),
the decay distribution in the angle $\theta$ is given by
\begin{eqnarray}
\frac{d\Gamma}{ds_{13} ds_{23} d\cos\theta} = C(s_{13}, s_{23}) \left\{
|\vec J|^2 (1 + \cos^2\theta) + \lambda_\gamma 2 \mbox{Im}[\hat n \cdot
(\vec J \times \vec J^*)] \cos\theta \right\}\,.
\end{eqnarray}

The second term in this decay distribution is sensitive to the photon polarization
parameter $\lambda_\gamma$. Its contribution can be isolated by
forming an up-down asymmetry with respect to the angle $\theta$. 
At each point $(s_{13},s_{23})$ in the Dalitz plot one may define an 
up-down asymmetry with respect to the $\hat e_z$ axis
\bea
\mathcal{A}(s_{13},s_{23}) \equiv \frac{1}{d\Gamma/(ds_{13}ds_{23})} 
\left(\int_0^{\pi/2} d\cos \theta
\frac{d\Gamma}{ds_{13}ds_{23} d\cos\theta} - 
\int_{\pi/2}^\pi d\cos \theta
\frac{d\Gamma}{ds_{13}ds_{23} d\cos\theta} \right) \,.
\eea

We have seen in (\ref{C'}) that for $K\pi\pi$ final states including a $\pi^0$ the overall
contribution from the two $K^*\pi$ intermediate states to $C_{1,2}$, which  
enters the definition of $\vec J$ in (\ref{J}), is antisymmetric under an exchange of 
the two pion momenta. Consequently the interference of the two $K^*\pi$ 
contributions, which for an intermediate $K_1(1400)$ is a dominant source for a 
photon up-down asymmetry 
in $B \to K\pi\pi^0\gamma$ (see next subsection), is antisymmetric 
under $s_{13} \leftrightarrow s_{23}$, and thus vanishes when being integrated 
over the entire Dalitz plot. For this reason one redefines a slightly modified 
integrated up-down asymmetry by multiplying the numerator with 
$\mbox{sgn}(s_{13}-s_{23})$ which is also antisymmetric in $(s_{13},s_{23})$,
\begin{eqnarray}
\tilde \mathcal{A} \equiv \frac{1}{\frac83 \langle |\vec J|^2 \rangle} 
2\lambda_\gamma \langle \mbox{sgn}(s_{13}-s_{23}) \mbox{Im}(\hat n \cdot 
(\vec J \times \vec J^*))\rangle = 
\frac34 \frac{\langle \mbox{sgn}(s_{13}-s_{23}) \mbox{Im}(\hat n \cdot 
(\vec J \times \vec J^*))\rangle}{\langle |\vec J|^2\rangle} \lambda_\gamma~.
\nonumber\\
\end{eqnarray}
The angular brackets denote integration over the Dalitz plot,
$\langle \cdots \rangle  = \int\cdots ds_{13} ds_{23}$.
This asymmetry may be formulated also as an up-down asymmetry with 
respect to an angle $\tilde \theta$ defined by 
$\cos\tilde\theta \equiv \mbox{sgn}(s_{13}-s_{23}) \cos\theta$, where 
$\tilde \theta $ is the angle between $\hat e_z$ and the normal to the plane
determined by $\vec p_{fast} \times \vec p_{slow}$~\cite{sign}.

\subsection{Three mechanisms for a photon asymmetry}

Given the expressions of $C_i(s_{13},s_{23})$ occurring in amplitudes for 
$K_1\to K\pi\pi$ decays and the experimental information about these amplitudes
 as described in Sec.\,4, we are now ready to calculate the photon up-down
 asymmetry with respect to the $K\pi\pi$ decay plane. 

As mentioned in the introduction, a nonzero up-down asymmetry which is odd under time-reversal requires two interfering amplitudes with a nonzero relative phase 
due to final state interactions. We identify three types of interference which 
involve such potential phases:

\begin{itemize}
\item (a) Interference of amplitudes for two $K^*\pi$ intermediate states.
Such interference, involving $K^{*0}\pi^+, K^{*+}\pi^0$ and $K^{*0}\pi^0, 
K^{*+}\pi^-$ in $K_1^+ \to K^0\pi^+\pi^0$ and $K_1^0\to K^+\pi^-\pi^o$ respectively, occurs only in decays involving a final neutral pion.
The amplitude for these $K_1^{+,0} \to K\pi\pi^0$ decays is given in Sec.\,4.2. 
The relevant strong phase originates in an overlap of two isospin-related 
Breit-Wigner $K^{*0}$ and $K^{*+}$ resonance bands in the Dalitz plot. The contribution of this interference to the asymmetry includes also interference
of $S$ and $D$ wave $K^*\pi$ amplitudes which depends on $\delta_{DS}^{(K^*\pi)}$
and vanishes for $\delta_{DS}^{(K^*\pi)}=0$.
We denote an asymmetry from interference of this kind by $\tilde{\cal A}_a$.
\item (b) Interference between $K^*\pi$ and $K\rho$ amplitudes. Such interference  
occurs  in all $K_1 \to K\pi\pi$ decays including both $K_1^+ \to K^+\pi^+\pi^-,
K_1^0\to K^0\pi^+\pi^-$ and $K_1^+ \to K^0\pi^+\pi^0,
K_1^0\to K^+\pi^-\pi^0$. This contribution to an asymmetry is affected by an 
overlap in the Dalitz plot of the $K^*$ and $\rho$ bands and 
depends on the two relative phases $\delta_{DS}^{(K^*\pi)}$ and $\alpha_S$.  
We denote this contribution to an asymmetry by $\tilde{\cal A}_b$.
\item (c) Interference of $S$ and $D$ wave amplitudes in $K_1 \to K^*\pi$. 
This kind of interference occurs in all four $K_1 \to K\pi\pi$ charged modes. 
Because of an assumed negligible $D$ wave amplitude in $K_1 \to K\rho$ due to very limited available phase space (in particular in $K_1(1270) \to K\rho$), we 
neglect a similar interference in these decays. 
The interference between $S$ and $D$ wave 
$K^*\pi$ amplitudes does not depend on overlapping bands in the Dalitz plot 
and on $\alpha_S$. 
The resulting asymmetry depends on $\alpha_S$ (through the asymmtry denominator) 
and on $\delta_{DS}^{(K^*\pi)}$ and vanishes for $\delta^{(K^*\pi)}_{DS}=0$.
This contribution to an asymmetry will be denoted $\tilde{\cal A}_c$.
\end{itemize}

Results will now be presented for up-down photon asymmetries with respect to 
the $K\pi\pi$ decay plane, which we calculate separately 
for decays involving $K_1(1400)$ and $K_1(1270)$ resonant states. 
In addition to total asymmetries we will 
present asymmetries due to interference of type (a) in decays involving a final 
neutral pion, and due to interference of types (b) and (c) for decays involving a 
$\pi^+\pi^-$ pair. We point out that the total asymmetry in the latter decays  is 
the sum $\tilde{\cal A}_{\rm total} = \tilde{\cal A}_b + \tilde{\cal A}_c$,  
in which $\tilde A_b$ and $\tilde A_c$ depend on both $\delta_{DS}^{(K^*\pi)}$
and $\alpha_S$.

\subsection{Photon asymmetry due to $B \to K_1(1400)\gamma$}
\subsubsection{$B^+\to K^0\pi^+\pi^0\gamma$ and $B^0\to K^+\pi^-\pi^0\gamma$}
\begin{table}[h]
\caption{Up-down photon asymmetry $\tilde{\cal A}$
in $B^+\to K^0\pi^+\pi^0\gamma$ from intermediate $K_1(1400)$. 
The asymmetry $\tilde{\cal A}_a$ neglects a contribution of a $\rho K$ 
amplitude as described in the text. For the total asymmetry we use 
$\alpha_S=40^\circ$, a value favored by the analysis of~\cite{Daum:1981hb}.}
\begin{center}
\begin{tabular}{c | c c c c c c c c}
\hline
$\delta^{(K^*\pi)}_{DS}$(degrees) & 
   0    & 45   & 90   & 135  & 180 & 225 & 270 & 315 \\
\hline
$\tilde{\cal A}_a$ & 
   0.30 & 0.21 & 0.14 & 0.14 & 0.19 & 0.28 & 0.34 & 0.35 \\
$\tilde{\cal A}_{\rm total}$ & 
   0.30 & 0.21 & 0.15 & 0.14 & 0.20 & 0.29 & 0.35 & 0.36  \\ 
\hline
\end{tabular}
\label{tab.asym1400(2)}
\end{center}
\end{table}
Table \ref{tab.asym1400(2)} shows total asymmetries and asymmetries of type (a)
calculated for a large range of phases $\delta_{DS}^{(K^*\pi)}$, assuming for the 
total asymmetry a value $\alpha_S=40^\circ$ favored by \cite{Daum:1981hb}.
We note that in decays involving a final state $\pi^0$ the total asymmetry 
is completely dominated by interference of type (a) of two amplitudes for two 
$K^*\pi$ intermediate states and is therefore practically independent on $\alpha_S$. 
This follows from the dominance of the $K^*\pi$ mode and the negligible  $K_1(1400)$
decay branching ratio into $K\rho$. The asymmetry $\tilde A_{\rm total}=0.30$ at 
$\delta_{DS}^{(K^*\pi)}=0$ is purely due to to an overlap of two equal strength (by isospin) 
Breit-Wigner $K^{*0}$ and $K^{*+}$ bands  in the Dalitz plot.
Using a value of $\delta_{DS}^{(K^*\pi)}$ 
around $260^\circ$, as indicated by the partial wave analysis performed in 
Ref.~\cite{Daum:1981hb}, one expects a slightly larger asymmetry of 
$34\%$~\cite{Gronau:2001ng,Gronau:2002rz}.
\subsubsection{$B^+ \to K^+\pi^+\pi^-\gamma$ and $B^0 \to K^0\pi^+\pi^-\gamma$.}
\begin{table}[h]
\caption{Up-down photon asymmetry $\tilde{\cal A}$ in 
$B^+\to K^+\pi^+\pi^-\gamma$ from intermediate $K_1(1400)$. 
Asymmetries $\tilde{\cal A}_b$ and $\tilde{\cal A}_c$ are defined in the text. 
The asymmetries are calculated for 
 $\alpha_S=40^\circ$, a value favored by the analysis in~\cite{Daum:1981hb}.}
\begin{center}
\begin{tabular}{c | c c c c c c c c}
\hline
$\delta^{(K^*\pi)}_{DS}$ (degrees) & 
 0 & 45 & 90 & 135 & 180 & 225 & 270 & 315 \\
\hline
$\tilde{\cal A}_b$ & 
0.00 & 0.00 & 0.00 & 0.00 & 0.01 & 0.01 & 0.01 & 0.00 \\
$\tilde{\cal A}_c$ & 
 0 & -0.07 & -0.10 & -0.07 & 0.0 & 0.07 &  0.10 & 0.07  \\
$\tilde{\cal A}_{\rm total}$ & 
 0 & -0.07 & -0.10 & -0.07 & 0.01 & 0.08 & 0.11 & 0.07 \\ 
\hline
\end{tabular}
\label{tab.asym1400(3)} 
\end{center}
\end{table}
Table \ref{tab.asym1400(3)} presents asymmetries of types (b) and (c) and 
total asymmetries for the same range of values of $\delta_{DS}^{(K^*\pi)}$ as in 
Table \ref{tab.asym1400(2)}, assuming $\alpha_S= 40^\circ$ as mentioned above and 
$|C_D^{(K^*\pi)}|/|C_S^{(K^*\pi)}| = 0.2$ (See Table \ref{tab.K1400}.)
The total asymmetry is seen to be dominated by terms of type (c) due to interference 
of $S$ and $D$ wave amplitudes in $K_1(1400) \to K^*\pi$, while terms of type (b) originating 
in interference between $K^*\pi$ and  $K\rho$ amplitudes are negligible. This can be traced
back to the very small $K\rho$ branching ratio of $K_1$ decay which is completely dominated
by $K^*\pi$. (See Table \ref{tab.K1400}.) While for arbitrary $\delta_{DS}^{(K^*\pi)}$ the total asymmetry may be positive or negative, it is predicted to be about $+10\%$ for 
$\delta_{DS}^{(K^*\pi)} \sim 260^\circ$ which is favored by the analysis in 
Ref.~\cite{Daum:1981hb}.
\subsection{Photon asymmetry due to $B \to K_1(1270)\gamma$}
\subsubsection{$B^+\to K^0\pi^+\pi^0\gamma$ and $B^0\to K^+\pi^-\pi^0\gamma$}
\begin{table}[h]
\caption{Up-down photon asymmetry in $B^+\to K^0\pi^+\pi^0\gamma$
from intermediate $K_1(1270)$. The asymmetry $\tilde{\cal A}_a$ neglects a contribution of a $\rho K$ 
amplitude as described in the text. For the total asymmetry we assume 
$\alpha_S=-40^\circ$, an approximate value obtained from 
Ref.~\cite{Guler:2010if} by assuming $S$ wave dominance of $K_1(1270)$ 
decays to $K^*\pi$ and $K\rho$.}
\begin{center}
\begin{tabular}{c | c c c c c c c c}
\hline
$\delta^{(K^*\pi)}_{DS}$(degrees) & 
   0 & 45 & 90 & 135 & 180 & 225 & 270 & 315 \\
\hline
$\tilde{\cal A}_a$ & 
   0.02 & -0.00 & -0.04 & -0.10 & -0.05 & 0.08 & 0.06 & 0.04 \\
$\tilde{\cal A}_{\rm total}$ & 
 -0.09 & -0.10 & -0.10 & -0.10 & -0.07 & -0.07 & -0.08 & -0.09  \\ 
\hline
\end{tabular}
\label{tab.asym1270(2)}
\end{center}
\end{table}
Table \ref{tab.asym1270(2)} shows total asymmetries and asymmetries of type (a) 
due to interference of amplitudes for two $K^*\pi$ intermediate states for 
$B^+\to K^0\pi^+\pi^0\gamma$ decays via $K_1(1270)$ as functions of 
$\delta_{DS}^{(K^*\pi)}$. The total asymmetry is predicted to lie in a narrow range 
between $-7\%$ and $-10\%$, considerably smaller than the corresponding asymmetry 
via $K_1(1400)$ given in Table \ref{tab.asym1400(2)}. While the latter was shown to 
be positive the former is negative. Unlike the situation we encountered with $K_1(1400)$, 
the total asymmetry via $K_1(1270)$ is not dominated by interference of type (a). 
This can be traced back to the small branching ratio of $K_1(1270)$ decay into $K^*\pi$ 
relative to its considerably larger decay rate into $\rho K$. 
\subsubsection{$B^+ \to K^+\pi^+\pi^-\gamma$ and $B^0 \to K^0\pi^+\pi^-\gamma$}
\begin{table}[h]
\caption{Up-down photon asymmetry in $B^+\to K^+\pi^+\pi^-\gamma$
from intermediate $K_1(1270)$ assuming $|C_D^{(K^*\pi)}|/|C_S^{(K^*\pi)}| = 1$. 
The asymmetries $\tilde{\cal A}_b$ and $\tilde{\cal A}_c$ are defined in the text. 
A full range of values for $\tilde{\cal A}_{\rm total}$ is calculated for selected 
values of $\delta_{DS}^{(K^*\pi)}$ and $\alpha_S$.}
\begin{center}
\begin{tabular}{c | c c c c c c c c}
\hline
$(\delta^{(K^*\pi)}_{DS}, \alpha_S)$ (degrees) & 
   (90,0) & (270,270) & (225,135) & (30,30) \\
\hline
$\tilde{\cal A}_b$ & 
   -0.05 & -0.08 & +0.12  & -0.05 \\
$\tilde{\cal A}_c$ & 
  -0.08 & +0.08 & +0.12  & -0.02 \\
$\tilde{\cal A}_{\rm total}$ & 
  -0.13 & +0.00 & +0.24 & -0.07 \\ 
\hline
\end{tabular}
\label{tab.asym1270(3)}
\end{center}
\end{table}
\begin{table}[h]
\caption{Up-down photon asymmetry in $B^+\to K^+\pi^+\pi^-\gamma$
from intermediate $K_1(1270)$ assuming $|C_D^{(K^*\pi)}|/|C_S^{(K^*\pi)}| = 0.2$. 
The asymmetries $\tilde{\cal A}_b$ and $\tilde{\cal A}_c$ are defined in the text. 
A full range of values for $\tilde{\cal A}_{\rm total}$ is calculated for selected 
values of $\delta_{DS}^{(K^*\pi)}$ and $\alpha_S$.}
\begin{center}
\begin{tabular}{c | c c c c c c c c}
\hline
$(\delta^{(K^*\pi)}_{DS}, \alpha_S)$ (degrees) & 
    (90,0) & (270,270) & (225,135) & (30,30) \\
\hline
$\tilde{\cal A}_b$ & 
   -0.05 & -0.08 & +0.09 & -0.04 \\
$\tilde{\cal A}_c$ & 
  -0.05 & +0.05 & +0.05  & -0.02 \\
$\tilde{\cal A}_{\rm total}$ & 
  -0.11 & -0.03 & +0.14 & -0.06 \\   
\hline
\end{tabular}
\label{tab.asym1270(3b)}
\end{center}
\end{table}
Table \ref{tab.asym1270(3)}  shows photon asymmetries 
calculated for $B^+\to K^+\pi^+\pi^-\gamma$ from intermediate $K_1(1270)$
assuming $|C_D^{(K^*\pi)}|/|C_S^{(K^*\pi)}| = 1$. In the absence of
 experimental information on $\delta_{DS}^{(K^*\pi)}$ and $\alpha_S$ we  
 varied these two phases over their entire range of $0^\circ - 360^\circ$ searching
 for an overall range of $\tilde{\cal A}_{\rm total}$. The asymmetries presented in 
 the table correspond to four cases: The largest positive and negative total 
 asymmetries, $+24\%$ and $-13\%$, a vanishing total asymmetry (obtained also
 for other values of the two phases) and a 
 fourth case involving arbitrarily chosen two phases of $30^\circ$ each. 
 In Table \ref{tab.asym1270(3b)} we present asymmetries assuming 
 $|C_D^{(K^*\pi)}|/|C_S^{(K^*\pi)}| = 0.2$, calculated for the same four pairs 
of phases ($\delta_{DS}^{(K^*\pi)},\alpha_S)$ as in Table \ref{tab.asym1270(3)}.
 We also found two extreme values of the total asymmetry, $+15\%$ and $-13\%$, 
 obtained for phases $(270^\circ,135^\circ)$ and $(90^\circ,315^\circ)$, respectively,
 and asymmetries $\le 1\%$ obtained for other values including a continuum range 
 $(90^\circ-135^\circ,90^\circ-135^\circ)$. 
  
We conclude that without further phase information about $\alpha_S$ and 
$\delta_{DS}^{(K^*\pi)}$ the total asymmetry can 
have any value ranging from $-13\%$ to $+24\%$. Typical contributions of 
asymmetries of types (b) and (c) have comparable magnitudes which may 
enhance or cancel each other in the total asymmetry. Comparing the entries in
Tables \ref{tab.asym1270(3)} and \ref{tab.asym1270(3b)} shows that for
certain phases the value of $|C_D^{(K^*\pi)}|/|C_S^{(K^*\pi)}|$ may have a 
significant effect on the photon asymmetry.

\section{Isospin symmetry in $B\to K\pi\pi\gamma$}
We have pointed out that the two pairs of strong decay amplitudes of 
$K_1^+ \to K\pi\pi$ and $K_1^0\to K\pi\pi$ in Eqs.\,(\ref{type3}) and 
(\ref{type2}), for final states related by isospin reflection 
$u \leftrightarrow d$, are equal in the isospin symmetry limit. 
Does a similar relation hold approximately for corresponding 
weak decay amplitudes $B^+ \to K\pi\pi\gamma$ and $B^0\to K\pi\pi\gamma$?

Isospin breaking in radiative $B$ meson decays has been studied in the literature
and found to be small. For two recent brief reviews including theoretical and 
experimental references see Ref.~\cite{Jung:2015yma,Paz:2017wow}.  
Isospin asymmetry at a level of $5\%$, consistent with zero at $2\sigma$, was 
measured by the Belle and Babar collaborations for 
$B\to K^*\gamma$~\cite{Olive:2016xmw,Nakao:2004th,Aubert:2009ak}, 
\beq
A_I(B\to K^*\gamma) \equiv \frac{\Gamma(B^0 \to K^{*0}\gamma) - 
\Gamma(B^+ \to K^{*+}\gamma)}{\Gamma(B^0 \to K^{*0}\gamma) + 
\Gamma(B^+ \to K^{*+}\gamma)} = 0.052 \pm 0.026~.
\eeq
Isospin breaking  in inclusive radiative decays $B \to X_s\gamma$ is expected to be further suppressed and has been measured at this level by Babar~\cite{Aubert:2005cua},
\beq
A_I(B \to X_s\gamma) \equiv   \frac{\Gamma(B^0 \to X^0_S\gamma) - 
\Gamma(B^+ \to X^+_s\gamma)}{\Gamma(B^0 \to X_s^0\gamma) + 
\Gamma(B^+ \to X^+_s\gamma)} = -0.006 \pm 0.058 \pm 0.009 \pm 0.024~.
\eeq

Thus one may assume that the following two approximate isospin equalities hold 
at a few percent level also for radiative decays to the $K_1$ resonances:
\bea\label{B+B0}
A(B^+ \to K_1^+\gamma \to K^+\pi^+\pi^-\gamma) & \approx &
A(B^0 \to K_1^0\gamma \to K^0\pi^-\pi^+\gamma)~,
\nonumber\\
A(B^+ \to K_1^+\gamma \to K^0\pi^+\pi^0\gamma) & \approx &
A(B^0 \to K_1^0\gamma \to K^+\pi^-\pi^0\gamma)~.
\eea
At this level of approximation one may therefore study the photon polarization by 
combining data for charged and neutral $B$ decays. This should double 
the statistics. One must pay some attention to the definition of an up-down 
asymmetry for these two pairs of processes by considering the isospin reflection, $u \leftrightarrow d$, which relates the final kaon and two pions in $B^+$ decays to corresponding final mesons in $B^0$ decays. 

\section{Conclusion}
In this paper we reexamined, updated and extended a suggestion made fifteen 
years ago to measure the photon polarization in $b \to s\gamma$ by observing in 
$B \to K\pi\pi\gamma$ an asymmetry of the photon with respect to the $K\pi\pi$ 
plane. Asymmetries were calculated for different charged final states due to 
intermediate $K_1(1400)$ and $K_1(1270)$ resonant states. Three 
interference mechanisms were identified playing different roles in decays involving
these two kaon resonances. 

\begin{itemize}
\item The situation is quite simple in decays via $K_1(1400)$, for which an upper bound 
${\cal B}(B \to K_1(1400)\gamma) < (1.2-1.5)\times 10^{-5}$ has been measured
using less than $20\%$ of the Belle total data sample involving $\pi^+\pi^-$. 
As $K_1(1400)$ is dominated by $K^*\pi$ decays, 
the total symmetry in decays involving a final state $\pi^0$ is large and positive
favoring values around $30\%$ from an overlap of two Breit-Wigner $K^*$ bands
of equal strength. The asymmetry in decays involving a final state 
$\pi^+\pi^-$ pair, dominated by interference of $S$ and $D$ wave amplitudes 
 in $K_1(1400) \to K^*\pi$, is considerably smaller favoring a value around $10\%$. 
 As these asymmetries show some dependence on the phase $\delta^{(K*\pi)}_{DS}$ 
 between $S$ and $D$ wave amplitudes in $K_1(1400) \to K^*\pi$ which has only been 
 measured in \cite{Daum:1981hb}, an independent measurement 
 of this phase in dedicated amplitude analyses of $B\to J/\psi(\psi') K\pi\pi$ decays would be 
 useful.
 \item The situation is considerably more involved in decays via $K_1(1270)$, for which a 
 branching ration ${\cal B}(B \to K_1(1270)\gamma) \sim 4\times 10^{-5}$ has been measured. 
 There are two reasons for this situation. First, the $K_1(1270)$ 
 decays more frequently to $K\rho$ than to $K^*\pi$, for which the branching ratio
 is only around $20\%$. Consequently, the total asymmetry in decays involving a final state 
 $\pi^0$ is not dominated by interference of two intermediate $K^*\pi$ states.
 A second reason for being unable to predict an asymmetry in decays involving an 
intermediate $K_1(1270)$ is lack of information about final state interaction phases in 
 its decays to $K^*\pi$ and $K\rho$. Assuming $S$ wave dominance of $K_1(1270)$ 
 decays to $K^*\pi$ and $K\rho$ an analysis in \cite{Guler:2010if} implies a value 
 $\alpha_S \sim -40^\circ$ for the relative phase between these two amplitudes. 
 Using this value of $\alpha_S$ the asymmetry is predicted to be negative and at 
 most $-10\%$ for a final state involving a $\pi^0$.

The situation in decays via $K_1(1270)$ involving a final state $\pi^+\pi^-$ pair is 
more uncertain because there is no information about the two relevant phases, 
$\alpha_S$ and $\delta_{DS}^{(K^*\pi)}$, and there exists only a crude measurement 
of $|C_D^{(K^*\pi)}|/|C_S^{(K^*\pi)}|$. Varying these phases over their entire range 
of $0^\circ - 360^\circ$ we calculated total asymmetries between $-13\%$ and $+24\%$, 
depending to some extent on the ratio of $D$ and $S$ amplitudes. The asymmetry 
obtains comparable contributions from interference of  $K^*\pi$ and $K\rho$ amplitudes and interference of $S$ and $D$ wave $K^*\pi$ amplitudes, which 
may act constructively or destructively with respect to one another. Major progress 
in predicting these asymmetries would be achieved by measuring the phases 
$\alpha_S$ and $\delta_{DS}^{(K^*\pi)}$ and improving the current measurement 
of the $D$ to $S$ ratio  in $K_1(1270)\to K^*\pi$. This could be achieved in 
dedicated amplitude analyses of $B \to J/\psi(\psi') K\pi\pi$ decays to be 
performed in the future  by the Belle II Collaboration at 
SuperKEKB~\cite{Abe:2010gxa,Aushev:2010bq}.
\end{itemize}

Finally, in order to increase statistics in studies of the photon polarization,
we suggest using approximate isospin symmetry (\ref{B+B0}) for combining in the 
same analysis $B\to K\pi\pi\gamma$ decays for charged and neutral $B$ mesons. 
So far the Belle collaboration used less than $20\%$ of their total data sample 
to obtain the branching ratio (\ref{BRK1(1270)}) for $B^+ \to K^+_1(1270)\gamma$ 
and the separate upper bounds (\ref{BRup(1400)}) on  $B^+ \to K^+_1(1400)\gamma$ and 
$B^0\to K^0_1(1400)\gamma$ for final states involving $\pi^+\pi^-$~\cite{Yang:2004as}. 
We urge the Belle collaboration to combine $B^+$ and $B^0$ decays when analyzing 
their full data sample for these decays, and to  study also final states including a $\pi^0$ 
in combined $B^+$ and $B^0$ decay samples.  

\medskip
We wish to thank Karim Trabelsi for asking very useful questions 
which motivated this work and Jonathan Rosner for helpful correspondence.

\end{document}